\RequirePackage{fix-cm}
\documentclass[twocolumn,epjc3]{svjour3}  
\RequirePackage{graphicx}
\RequirePackage{color}
\journalname{Eur. Phys. J. C}
\begin{document}

\title{ Reply to ``Comment on Comment on Linear confinement of a scalar particle in a G\"{o}del-type space-time''
}

\author{Faizuddin Ahmed \thanksref{e1,addr1}
}

\thankstext{e1}{e-mail: faizuddinahmed15@gmail.com}

\institute{Ajmal College of Arts and Science, Dhubri-783324, Assam, India \label {addr1}
}

\date{Received: date / Accepted: date}

\maketitle

\begin{abstract}

A hypothetical energy quantization condition is imposed in \cite{Neto} (Francisco A. Cruz Neto {\it et al.}, arXiv : 1910.11701 [gr-qc]) in order to show consistent of the result with those in \cite{Carvalo} but fail or unable to show consistency with those in \cite{Vitoria}.

\end{abstract}

In a recent paper, Francisco A. Cruz Neto {\it et al.} \cite{Neto} have written down the radial wave-equation of the KG-equation in the Som-Raychaudhuri space-time subject to a linear scalar potential and with a suitable ansatz they have obtained expressions (1)--(13) with the recurrence condition (14) for $\theta \neq 0$ case. The recurrence condition (14) (symbols are same as in \cite{Neto}) is given by
\begin{eqnarray} 
A_{j+2}&=&\frac{\theta}{2}\,(2\,j+3+\frac{2\,|l|}{\alpha})\,A_{j+1}\nonumber\\
&&-(j+1)\,(j+1+\frac{2\,|l|}{\alpha})\,(\bigtriangleup-2\,j)\,A_j,
\label{1}
\end{eqnarray}
where $\bigtriangleup=\tau-2-\frac{2\,|l|}{\alpha}$. From the recurrence relation (1), one can see that the solution $H$ becomes a polynomial of degree $n$ if and only if $\bigtriangleup=2\,n$ ($n=0,1,2,.....$) and $A_{n+1}=0$ which implies $A_{n+2}=0$. 

Afterwards, without a recurrence relation the authors in \cite{Neto} imposed that if $\theta=0$, the solution $H$ becomes a polynomial of degree $n$ if and only if
\begin{equation}
\bigtriangleup=4\,n\quad (n=0,1,2,3,......).
\label{2}
\end{equation}
The above energy quantization condition is completely a hypothetical one. The authors \cite{Neto} very cleverly escaped not to write down a recurrence relation for $\theta=0$ case to obtain the solution $H$ a polynomial of degree $n$ provided the above condition (\ref{2}) holds. From where the above condition (\ref{2}) comes which gives the solution $H$ a polynomial of degree $n$ not metioned at all. 

Furthermore, for $\theta=0$ case the authors have written a condition (17) by $\beta-2-\frac{2\,|l|}{\alpha}=24\,n$ and obtain the following energy eigenvalues 
\begin{eqnarray}
E_{n,l}&=&(2\,n+1+\frac{l}{\alpha}+\frac{|l|}{\alpha})\,\Omega\nonumber\\
&&+\sqrt{(2\,n+1+\frac{l}{\alpha}+\frac{|l|}{\alpha})^2\,\Omega^2+M^2+k^2},
\label{3}
\end{eqnarray}
where $n=0,1,2,3,.....$. The energy eigenvalues (\ref{3}) is similar to the result obtained in \cite{Carvalo} (see Eq. (14) in \cite{Carvalo}). Interestingly, they failed to compare their result (\ref{3}) with those in \cite{Vitoria} given by (see Eq. (23) in \cite{Vitoria})
\begin{eqnarray}
E_{n,l}&=&(n+1+\frac{l}{\alpha}+\frac{|l|}{\alpha})\,\Omega\nonumber\\
&&+\sqrt{(n+1+\frac{l}{\alpha}+\frac{|l|}{\alpha})^2\,\Omega^2+M^2+k^2}.
\label{4}
\end{eqnarray}
The above eigenvalues (\ref{4}) is not consistent with the result obtained in \cite{Carvalo} the issue that was addressed in \cite{EPJC}. There is no other intention except the eigenvalues (\ref{4}) correctly obtained in \cite{EPJC}.

We see that a hypothetical condition (\ref{2}) is imposed to claim the solution $H$ a polynomial degree $n$ without a recurrence condition for $\theta=0$ case. The authors obtained the energy eigenvalues (\ref{3}) which is consistent with the result in Ref. \cite{Carvalo} (see Eq. (14) in Ref. \cite{Carvalo}) but they are unable to show consistency or compare with those in Ref. \cite{Vitoria} (see Eq. (23) in Ref. \cite{Vitoria}).

\end{document}